\documentclass[preprint,12pt]{JASA}
	
	\usepackage{amsmath}
	\usepackage{graphicx}
	\usepackage{enumerate}
	\usepackage{natbib} 
	\usepackage{url} 
	\usepackage{multirow}
	\usepackage{caption}

\renewcommand{\linenumbers}{}
	
\begin{document}

\title{A Bayesian Modified Ising Model for Identifying Spatially Variable Genes from Spatial Transcriptomics Data}

\author{Xi Jiang}\affiliation{Department of Statistical Science, Southern Methodist University, 3215 Daniel Avenue 
Dallas, TX 75275, United States}
\affiliation{Department of Population and Data Sciences, The University of Texas Southwestern Medical Center, Dallas, TX 75390, United States}

\author{Qiwei Li}\email{To whom correspondence should be addressed. Emails: Qiwei.Li@UTDallas.edu; Guanghua.Xiao@UTSouthwestern.edu} \affiliation{
Department of Mathematical Sciences,
The University of Texas at Dallas, 
800 W Campell Rd, Richardson, TX 75080, United States}
\author{Guanghua Xiao} \email{To whom correspondence should be addressed. Emails: Qiwei.Li@UTDallas.edu; Guanghua.Xiao@UTSouthwestern.edu} \affiliation{Department of Population and Data Sciences, The University of Texas Southwestern Medical Center, Dallas, TX 75390, United States}


\begin{abstract}
\textbf{Abstract} \\
A recent technology breakthrough in spatial molecular profiling has enabled the comprehensive molecular characterizations of single cells while preserving spatial information. It provides new opportunities to delineate how cells from different origins form tissues with distinctive structures and functions. One immediate question in spatial molecular profiling data analysis is to identify genes whose expressions exhibit spatially correlated patterns, called spatially variable genes. Most current methods to identify spatially variable genes are built upon the geostatistical model with Gaussian process to capture the spatial patterns, which rely on ad hoc kernels that could limit the models’ ability to identify complex spatial patterns. In order to overcome this challenge and capture more types of spatial patterns, we introduce a Bayesian approach to identify spatially variable genes via a modified Ising model.  The key idea is to use the energy interaction parameter of the Ising model to characterize spatial expression patterns. We use auxiliary variable Markov chain Monte Carlo algorithms to sample from the posterior distribution with an intractable normalizing constant in the model. Simulation studies using both simulated and synthetic data showed that the energy-based modeling approach led to higher accuracy in detecting spatially variable genes than those kernel-based methods. When applied to two real spatial transcriptomics datasets, the proposed method discovered novel spatial patterns that shed light on the biological mechanisms. In summary, the proposed method presents a new perspective for analyzing spatial transcriptomics data.\\
\textit{Keywords}: Spatial molecular profiling; Multitype point pattern; Spatial correlation, Dichotomization; Double Metropolis-Hastings
\end{abstract}

\maketitle

 \section{Introduction}\label{sec1}

Cellular and molecular spatial organizations play essential roles in biological functions. In recent years, spatial molecular profiling (SMP) techniques have made significant breakthroughs, which enable transcriptome measurement in high spatial resolution \citep{zhang2020spatial}. Gene expression profiling approaches are no longer limited tissue-dissociation, which led to the loss of spatial context of the measured gene expression \citep{femino1998visualization}. Sequencing-based spatial molecular profiling platforms, such as spatial transcriptomics (ST) \citep{staahl2016visualization} and the improved Visium platform, use spatial barcodes to capture RNA molecules and then synthesize and sequence their complementary DNA molecules. Through this technology, the expression levels of thousands of genes can be measured across hundreds of spatial locations, namely spots. Those spots are usually arrayed on a two-dimensional grid. Particularly, ST and Visium spots are arranged on the square and triangular lattices, respectively.

SMP techniques enable researchers to study the gene expressions together with their spatial and morphological contexts, which provide new opportunities to advance our understanding of both cellular and molecular spatial organizations \citep{crosetto2015spatially}, and their relationships with diseases \citep{shah2018dynamics}. Many new questions can be explored with the emerging SMP techniques. One of the most immediate ones is to identify genes whose expressions exhibit spatially correlated patterns, referred to spatially variable (SV) genes. The study of spatial patterns in gene expression could reveal significant insights into many aspects, such as embryo development \citep{satija2015spatial}, tumor progression \citep{de2014spatial}, and the clinical impact of intra-tumor heterogeneity \citep{bedard2013tumour}. 

Several methods have been developed in recent years to address the above fundamental question in ST studies. Trendsceek \citep{edsgard2018identification} is based on marked point processes, and it is computationally intensive \citep{dries2021giotto} and has noticeably unsatisfying performance \citep{sun2020statistical}. BinSpect \citep{dries2021giotto} is an easy and fast computational method based on statistical enrichment of spatial network neighbors after binarizing gene expression levels. Most model-based analyses, such as SPARK \citep{sun2020statistical}, SpatialDE \citep{svensson2018spatialde}, and BOOST-GP \citep{li2020bayesian}, are built upon the geostatistical model with Gaussian process (GP), where a kernel must be selected with caution. Among these methods, SpatialDE transforms the measured counts at different locations into normalized data before analysis, SPARK uses a Poisson distribution to model the count data directly, and BOOST-GP accounts for the zero-inflation and mean-variance relationship existed in raw counts under a Bayesian framework. However, the GP-based models rely on ad hoc kernels that limit the models’ ability to identify complex spatial patterns. Furthermore, none of the existing approaches take advantage of the additional spatial structure of the ST data; that is, the gene expression levels are measured on a lattice grid. 

To enable the model to identify complex spatial patterns, we developed a novel approach named Bayesian mOdeling Of Spatial Transcriptomics data via a Modified Ising model (BOOST-MI) to identify SV genes in ST studies. It makes computation efficient by taking advantage of the fact that the ST experiments measure gene expression on a lattice. Ahead of fitting BOOST-MI, we need to normalize sequence count data to relative gene expression levels and dichotomize relative expression levels to a binary spatial pattern. Then, BOOST-MI characterizes the binary spatial pattern via inferring the Ising model interaction parameter under a Bayesian framework. The double Metropolis-Hastings (DMH) algorithm \citep{liang2010double} is used to sample from the posterior distribution with an intractable normalizing constant in the Ising model. Compared with other existing approaches, BOOST-MI tests the interaction energy parameter in a modified Ising model, which is able to characterize a broader type of spatial patterns than kernel-based modeling approaches. In addition, the proposed method enriches the inference via incorporating priors if necessary and naturally quantifying uncertainties under a Bayesian framework. We demonstrated the advantages of BOOST-MI in a comprehensive simulation study using both simulated data with various spatial patterns and zero-inflation settings and synthetic data from real ST experiments. The proposed model showed an outstanding performance compared to the alternatives. Finally, when applied to two real ST datasets, the proposed method discovered novel spatial patterns that shed light on the biological mechanisms.

The remainder of the paper is organized as follows. Section 2 introduces the two main components of the data preparation for BOOST-MI: sequence count data normalization and relative expression level dichotomization. In Sections 3 and 4, we introduce the proposed modeling framework and describe the Markov chain Monte Carlo (MCMC) algorithm and the resulting posterior inference. In Section 5, we compare BOOST-MI with existing approaches on simulated and two real ST datasets. Section 6 concludes the paper and discusses future research directions.

\section{Data Preparation}\label{sec2}

In this section, we introduce the two data preprocessing steps (see the dashed-line block in Figure \ref{flow}) before fitting BOOST-MI to identify SV genes. We first summarize the ST data notations. Let an $ n\times p$ matrix $\boldsymbol{Y}$ denote the gene expression count table (i.e. the molecular profile). Each entry $y_{ij} \in \mathbb{N}, i=1,\ldots,n,j=1,\ldots,p$ is the read count for gene $j$ collected at spot $i$. We use an $n\times 2$ matrix $\boldsymbol{T}$ to represent the geospatial profile, where each row $\boldsymbol{t}_i = (t_{i1}, t_{i2})$ indicates the spot location on the two-dimensional space. All non-boundary spots have the same neighborhood structure. Suppose the gene expression levels are measured on a $L$-by-$W$ square lattice grid, the coordinate of each spot $i$ can be written as $(t_{i1}=l,t_{i2}=w), l,=1,\ldots,L,w=1,\ldots,W$. If it is not a boundary spot, then its four neighboring spots are at locations $(l\pm1,w)$ and $(l,w\pm1)$.

Normalization is critical to sequence count data analysis. To counteract various artifacts and bias due to biological and technical reasons, we convert each read count to its relative gene expression level, denoted by $\tilde{y}_{ij} = y_{ij}/s_i$, where $s_i$ is the size factor of sample $i$, capturing all nuisance effects. The most straightforward way is to set $s_i\propto Y_i=\sum_{j=1}^py_{ij}$, i.e. the total number of counts across all genes in each sample (known as sequencing depth or library size), combined with some constraint, such as $\prod_{i=1}^ns_i=1$. Note that SPARK suggests this normalization, namely total sum scaling (TSS). We could also consider other estimations on $s_i$'s, which have been introduced for mitigating the influence of extremely low and high counts when analyzing bulk RNA-seq data, such as upper-quartiles (Q75) \citep{bullard2010evaluation}, relative log expression (RLE) \citep{anders2010differential}, and weighted trimmed mean by M-values (TMM) \citep{robinson2010scaling}. In addition to the above normalization method based on size factor estimation, we could use the method based on variance-stabilizing transformation (VST) (detailed in Section S1 in the supplement). A following sensitivity analysis (see Figure S1 in the supplement) indicates that 1) the methods based on size factor estimation noticeably outperformed the VST-based methods; 2) the performance based on TSS was not significantly different from Q75, RLE, and TMM. Thus, we suggest using TSS as the default setting for the sake of simplicity.

We then denoise the relative expression levels by partitioning all spots into two groups. This step outputs the suitable data type required in the subsequent analysis and makes our approach more robust in the face of over-dispersion and zero-inflation, which are the two important characteristics of ST data. For each gene, we introduce a binary vector $\boldsymbol{p}_j=(p_{1j},\ldots,p_{nj})$ to represent the dichotomization result based on its relative expression levels $(\tilde{y}_{1j},\ldots,\tilde{y}_{nj})$, with $p_{ij}=1$ indicating gene $j$ is highly expressed at spot $i$ and $p_{ij}=0$ otherwise. Since there is no consensus on the dichotomization of spots based on either absolute or relative expression level, we propose to estimate $\boldsymbol{p}_j$ via fitting a two-component Gaussian mixture model (GMM), $\tilde{y}_{ij}|p_{ij},\boldsymbol{\mu}_j,\boldsymbol{\sigma}_j^2\sim (1-p_{ij})\text{N}(\mu_{j0},\sigma_{j0}^2)+p_{ij}\text{N}(\mu_{j1},\sigma_{j1}^2)$, subjecting to $\mu_{j0}<\mu_{j1}$. Here $\boldsymbol{\mu}_j=(\mu_{j0},\mu_{j1})$ and $\boldsymbol{\sigma}_j^2=(\sigma_{j0}^2,\sigma_{j1}^2)$ are the group means and variances that need to be estimated. In addition to the model-based clustering, we also consider $k$-means, which is implicitly based on pairwise distances between relative expression levels. Section S2 in the supplement provides details about the GMM and $k$-means implementations ($k=2$). A following sensitivity analysis (see Figure S2 in the supplement) indicates that BOOST-MI performed equally well between the two choices.

\section{Models}
In this section, we first review the Ising model and then introduce a modified Ising model with external fields to identify SV genes. The schematic diagram is shown in Figure \ref{flow}.

\subsection{A brief review of the Ising model}
\begin{figure}[!p]
    \centering
    \includegraphics[width = 0.8\textwidth]{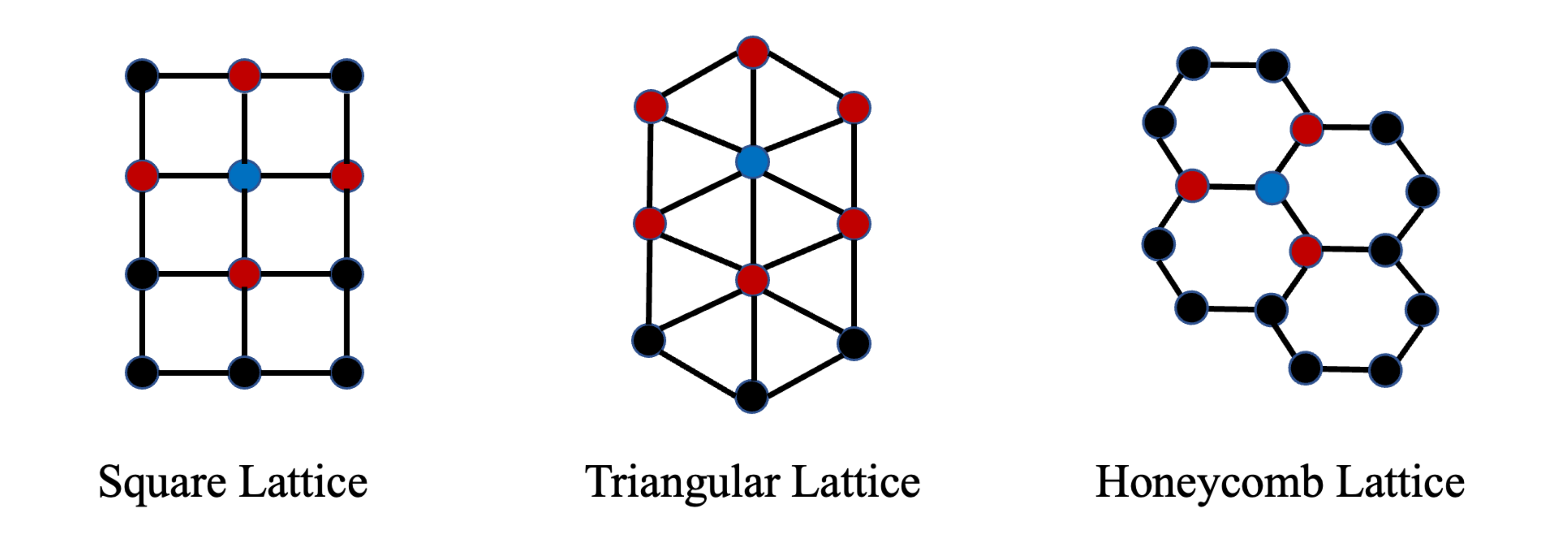}
    \caption{Common types of two-dimensional lattices. Red spots are the neighbors of the blue spot, while black spots are not.}
    \label{lattice}
\end{figure}

The Ising model, first introduced by \cite{lenz1920beitrvsge} and used in statistical mechanics, is a model of interacting binary states on a crystalline lattice \citep{cipra1987introduction}. Let $G = (V,E)$ denote a graph with a finite set of vertices $V$ and a set of edges $E$. In statistics, the Ising model is considered as an undirected graph such that each vertex is geometrically regular assigned on a lattice and each edge is of the same length \citep{tucker1994applied}. There are three typical types of two-dimensional lattices: square, triangular, and honeycomb (see Figure \ref{lattice}). The former two are a special case of Bravais lattices, which can be defined as an infinite array of discrete points generated by a set of discrete translation operations \citep{ashcroft1976solid}. ST and Visium (an improved ST platform) spots are arranged on square and triangular lattices, where each non-boundary spot has four and six neighbors, respectively. Every vertex will be assigned a binary state (also known as a spin). If there are three or more states, the model is known as the Potts model. Because the spots are assigned with different spins and react with their neighbors' spins, there exists a measurement of overall energy, named Hamiltonian,
\begin{eqnarray}\label{H0}
	H(\boldsymbol{p}_j|\theta_j)= -\theta_j \sum_{i \sim i'}I(p_{ij} \neq p_{i'j}),
\end{eqnarray}
where $i\sim i'$ denotes the collection of all neighboring spot pairs, $\theta_j$ denotes the interaction energy between highly and lowly expressed spots for gene $j$, and $I(\cdot)$ denotes the indicator function. According to the Hammersley–Clifford theorem \citep{clifford1990markov}, we can write the probability of observing a particular configuration of $\boldsymbol{p}_j$ as $\text{Pr}(\boldsymbol{p}_j|\theta_j)= {\exp(-H(\boldsymbol{p}_j|\theta_j))}/{\sum_{\boldsymbol{p}'\in\mathcal{P}} \exp(-H(\boldsymbol{p}'|\theta_j))}$, where $\mathcal{P}$ denotes the set of all configurations of spins. An exact evaluation of the normalizing constant (i.e. the denominator of the above equation) requires us to sum over the entire space, consisting of $2^{n}$ configurations. Thus, $\text{Pr}(\boldsymbol{p}_j|\theta_j)$ is intractable even for a small lattice size.

\subsection{A proposal of using a modified Ising model to identify SV genes}
Although Ising models have a wide range of applications in many areas, they are often examined assuming that the underlying abundance of different spins is equivalent. In the context of statistical mechanics, we can view the model as a system of interacting particles in the absence of an external field. However, we found that low-expression spots are usually much higher than high-expression spots due to excessive zeros presented in the ST data. For instance, the averaged proportions of low-expression spots in the two processed real ST datasets analyzed in the paper were $78.6\%$ and $86.8\%$, respectively (see Figure S3 in the supplement). Ignoring this feature could decrease the identification accuracy due to the inclusion of more false positives (see Figure \ref{simu_auc}).

\begin{figure}[!p]
    \centering
    \includegraphics[width = \textwidth]{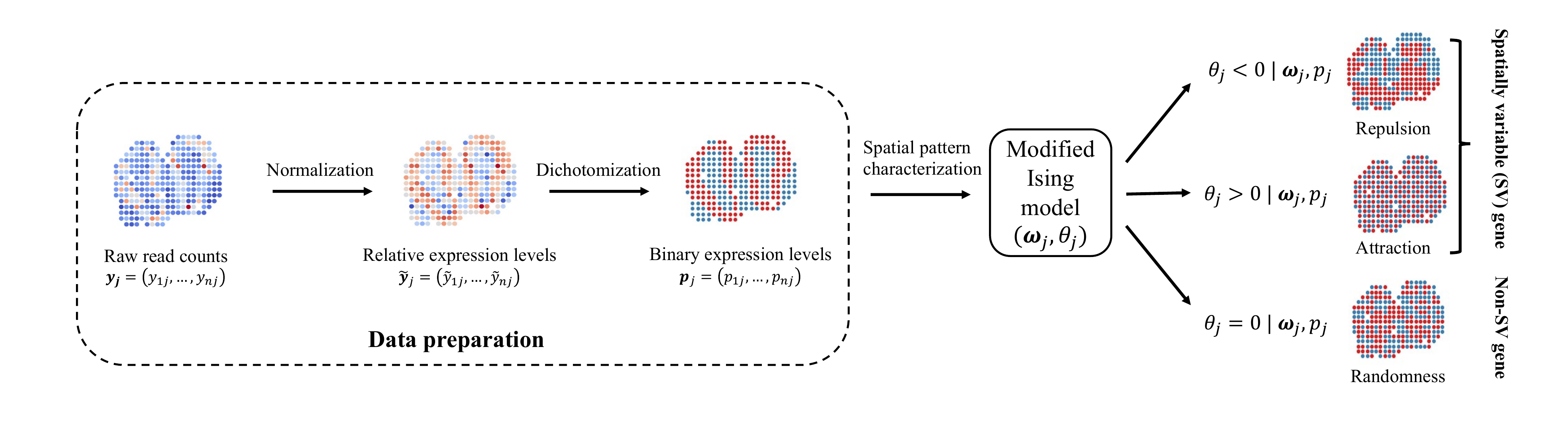}
    \caption{The schematic diagram of the proposed BOOST-MI model}
    \label{flow}
\end{figure}

To remedy the oversimplified Ising model, we add an external force in the Hamiltonian,
\begin{eqnarray}\label{H}
	H(\boldsymbol{p}_j|\boldsymbol{\omega}_j,\theta_j)= -\theta_j\sum_{i \sim i'} I(p_{ij} \neq p_{i'j})-\left[\omega_{j0}\sum_{i}I(p_{ij}=0)+\omega_{j1}\sum_{i}I(p_{ij}=1)\right],
\end{eqnarray}
where $\boldsymbol{\omega}_j=(\omega_{j0},\omega_{j1})$ and $\theta_j$ represent the first and second-order intensities. The first term is proportional to the number of neighboring spot pairs with different spins, while the remaining part (in the square bracket) can be viewed as the weighted average of the numbers of spots with different spins. As the energy function in Equation (\ref{H}) is still locally defined, we can write the joint probability on $\boldsymbol{p}_j$, up to its normalizing constant, as
\begin{eqnarray}\label{full_lklh}
	\begin{split}
	\text{Pr}(\boldsymbol{p}_j|\boldsymbol{\omega}_j,\theta_j)&\propto {\exp(-H(\boldsymbol{p}_j|\boldsymbol{\omega}_j,\theta_j))}\\
	&=\exp\left(\theta_j\sum_{i \sim i'} I(p_{ij} \neq p_{i'j})+\omega_{j0}\sum_{i}I(p_{ij}=0)+\omega_{j1}\sum_{i}I(p_{ij}=1)\right),
	\end{split}
\end{eqnarray}
which serves as the full data likelihood of the proposed BOOST-MI model. To interpret $\boldsymbol{\omega}_j$ and $\theta_j$, we give the conditional probability of observing a high-expression level of gene $j$ at spot $i$,
\begin{eqnarray}\label{lklh}
	\text{Pr}\left(p_{ij}=1|\cdot\right)\propto\exp\left(\omega_{j1}+\theta_j \sum_{i'\in\text{Nei}(i)}I(p_{i'j}=0)\right),
\end{eqnarray}
where $\text{Nei}(i)$ denotes the set of all neighbors to spot $i$. Equation (\ref{lklh}) is essentially a logistic regression, and hence the parameters $\boldsymbol{\omega}_j$ and $\theta_j$ can be interpreted in terms of conditional odds ratios in general. According to Equation (\ref{lklh}), if $\theta_j=0$, then $\text{Pr}\left(p_{ij}=1|\cdot\right)=\exp(\omega_{j1})/\exp(\omega_{j0}+\omega_{j1})$, implying that each dichotomized expression level $p_{ij}$ is independently and identically sampled from a Bernoulli distribution. Thus, no spatial pattern should exhibit, and gene $j$ is a non-SV gene. The underlying abundance of low and high-expression levels in $\boldsymbol{p}_j$ are characterized by $\exp(\omega_{j0})/(\exp(\omega_{j0})+\exp(\omega_{j1}))$ and $\exp(\omega_{j1})/(\exp(\omega_{j0})+\exp(\omega_{j1}))$, respectively. Fixing $\boldsymbol{\omega}_j$, Equation (\ref{lklh}) reveals that the smaller the $\theta_j$ (i.e. $\theta_j\rightarrow-\infty$), the more likely the dichotomized expression level at any spot is concordant with the majority of its neighboring spots' levels, resulting in a repulsion pattern (i.e. the clustering of spots with the same dichotomized expression level). In contrast, when $\theta_j$ takes a large positive value, we expect an attraction pattern; that is, the exhibition of clustering among spots with different dichotomized expression levels. Thus, the spatial correlation between the low and high-expression levels of gene $j$ can be quantified by $\theta_j$. Figure \ref{flow} shows the three typical patterns in terms of $\theta_j$ conditional on $\omega_{j0}=\omega_{j1}$. 

In conclusion, BOOST-MI uses $\theta_j$ to characterize the binary spatial pattern defined by the dichotomized gene expression levels $\boldsymbol{p}_j$ on the lattice grid $\boldsymbol{T}$. It is noteworthy that the kernel-based methods, such as SPARK, SpatialDE, and BOOST-GP, are only able to identify SV genes with a small subset of repulsion patterns defined by the selected kernel, while BOOST-MI accounts for all attraction and repulsion patterns. 

To complete the model specification, we impose $\theta_{j}\sim\text{N}(0,\sigma_\theta^2)$. As for the first-order intensity $\boldsymbol{\omega}_j$, we notice that an identifiability problem arises from Equation(\ref{lklh}). For example, adding a nonzero constant $c$ into $\omega_{jk},k=0,1$ does not change the conditional probability $\text{Pr}\left(p_{ij}=k|\cdot\right)$. Thus, we force $\omega_{j1}=1$ and set a normal prior on $\omega_{j0}\sim\text{N}(1, \sigma_\omega^2)$.

\section{Model Fitting}
In this section, we describe the MCMC algorithm for model fitting and the posterior inference. Our inferential strategy allows for simultaneously estimating the first-order intensity $\boldsymbol{\omega}_j$, which reveals the underlying abundance of the low and high-expression levels of gene $j$, and the second-order intensity $\theta_j$ (also known as the interaction parameter), which captures the spatial correlation between the low and high-expression levels. We give the details of our MCMC algorithm and the resulting posterior inference as below. Note that each gene is tested independently by BOOST-MI.

\subsection{MCMC algorithm}
The full data likelihood function is given in Equation (\ref{full_lklh}), which involves an intractable normalizing constant $C(\boldsymbol{\omega}_j, \theta_j) = \sum_{\boldsymbol{p}'\in\mathcal{P}} \exp(-H(\boldsymbol{p}'|\boldsymbol{\omega}_j,\theta_j))$. Taking the two real ST datasets analyzed in the paper as examples, it needs to sum over $2^{250}\approx1.8\times10^{75}$ and $2^{260}\approx1.9\times10^{78}$ elements, respectively. This makes the Metropolis–Hastings (MH) algorithm infeasible in practice. To overcome this challenge, we employ the double MH (DMH) algorithm \citep{liang2010double} to estimate both $\boldsymbol{\omega}_j$ and $\theta_j$ for each gene. The DMH is an asymptotic algorithm, which has been shown to produce accurate results by various spatial models \citep{li2019bayesian2,li2019bayesian}. Unlike other auxiliary variable MCMC algorithms \citep{moller2006efficient,murray2012mcmc} that also aim to have the normalizing constant ratio canceled, it is more efficient because it does not require drawing the auxiliary variables from a perfect sampler, which is usually computationally expensive.

To update $\theta_j$ within each iteration, we first simulate a new sample $\theta_j'$ from $\pi(\theta_j)$ using the MH algorithm starting with $\theta_j$. Then, we generate an auxiliary variable $\boldsymbol{p}_j'$ through $m$ MH updates starting with the current state $\boldsymbol{p}_j$ based on the new value $\theta_j'$ and accept it with probability $\text{min}(1, R_{\theta_j})$, where $R_{\theta_j}=\frac{\text{Pr}(\boldsymbol{p}_j'| \boldsymbol{\omega}_j, \theta_j)\text{Pr}(\boldsymbol{p}_j| \boldsymbol{\omega}_j ,\theta_j' )}{\text{Pr}(\boldsymbol{p}_j| \boldsymbol{\omega}_j, \theta_j)\text{Pr}(\boldsymbol{p}_j'| \boldsymbol{\omega}_j, \theta_j')} \frac{\pi(\theta_j')}{\pi(\theta_j)}$. If the auxiliary variable $\boldsymbol{p}_j'$ is accepted, we update $\theta_j$ to $\theta_j'$; otherwise, we keep the value of $\theta_j$. No improvement in the performance was noticed beyond $m=5$ in both simulation and application studies of this paper.

Just as in updating $\theta_j$, we use the DMH algorithm to update $\omega_{j0}$. Specifically, we first simulate a new sample $\boldsymbol{\omega}_j' = (\omega_{j0}', 1)$ from $\pi(\omega_{j0})$ using the MH algorithm starting with $\omega_{j0}$. Then, we generate an auxiliary variable $\boldsymbol{p}_j'$ through $m$ MH updates starting with the current state $\boldsymbol{p}_j$ based on the new value $\boldsymbol{\omega}_j'$ and accept it with probability $\text{min}(1, R_{\omega_j})$, where $R_{\omega_j} = \frac{\text{Pr}(\boldsymbol{p}_j'| \boldsymbol{\omega}_j, \theta_j)\text{Pr}(\boldsymbol{p}_j| \boldsymbol{\omega}_j', \theta_j)}{\text{Pr}(\boldsymbol{p}_j| \boldsymbol{\omega}_j, \theta_j)\text{Pr}(\boldsymbol{p}_j'| \boldsymbol{\omega}_j', \theta_j)}\frac{\pi(\omega_{j0}')}{\pi(\omega_{j0})}$. If the auxiliary variable $\boldsymbol{p}_j'$ is accepted, we update $\omega_{j0}$ to $\omega_{j0} '$; otherwise, the value of $\omega_{j0}$ remains the same.

\subsection{Posterior inference}
Our primary interest lies in the identification of SV genes via making inferences on the interaction parameter $\theta_j$. We obtain the posterior inference by post-processing of the MCMC samples after burn-in. To validate if gene $j$ exhibits a repulsion pattern, we set the null and alternative hypotheses as $\mathcal{M}_0:\theta_j\ge0$ and $\mathcal{M}_1:\theta_j<0$; while testing an attraction pattern, we set $\mathcal{M}_0:\theta_j\le0$ and $\mathcal{M}_1:\theta_j>0$. We could select the model via calculating the Bayes factor (BF) in favor of $\mathcal{M}_1$ over $\mathcal{M}_0$, which is defined as the ratio of posterior odds to prior odds,
\begin{eqnarray}\label{BF}
\text{BF}_j=\frac{\frac{\text{Pr}( \mathcal{M}_1 | \ \boldsymbol{p}_{\cdot j})}{\text{Pr}( \mathcal{M}_0 | \boldsymbol{p}_{\cdot j})}}{ \frac{\text{Pr}(\mathcal{M}_1)}{\text{Pr}(\mathcal{M}_0)}}\approx\begin{cases}\begin{array}{ll}
\frac{\sum_{u=1}^{U}I\left(\theta^{(u)}_j < 0\right)/U}{\sum_{u=1}^{U}I\left(\theta^{(u)}_j \ge 0\right)/U} & \text{for repulsion}\\
\frac{\sum_{u=1}^{U}I\left(\theta^{(u)}_j > 0\right)/U}{\sum_{u=1}^{U}I\left(\theta^{(u)}_j \le 0\right)/U} & \text{for attraction}
\end{array}\end{cases},
\end{eqnarray}
where the prior odds cancel out as we choose a normal prior on $\theta_j$ centered at zero, and the posterior odds can be approximated using the MCMC samples $\{\theta_j^{(1)},\ldots,\theta_j^{(U)}\}$. Here $U$ denotes the total number of MCMC iterations after burn-in. The larger the $\text{BF}_j$, the more likely gene $j$ is an SV gene, integrating over the uncertainty in all model parameters. We suggest choosing the BF threshold based on the scale for interpretation \citep{kass1995bayes}.

\section{Results}
\label{sec3}

\subsection{Simulation}
We performed a series of simulation studies to evaluate the performance of BOOST-MI and compared it with that of four existing methods: SpatialDE, SPARK,  BOOST-GP, and BinSpect. Because of the poor performance of Trendsceek reported in most literature \citep{sun2020statistical, dries2021giotto}, we did not include it here. In addition, we fitted the classic Ising model (with the Hamiltonian defined in Equation (\ref{H0})) under the same Bayesian framework.

\begin{figure}[!p]
	\centering
	\includegraphics[width =0.85\textwidth]{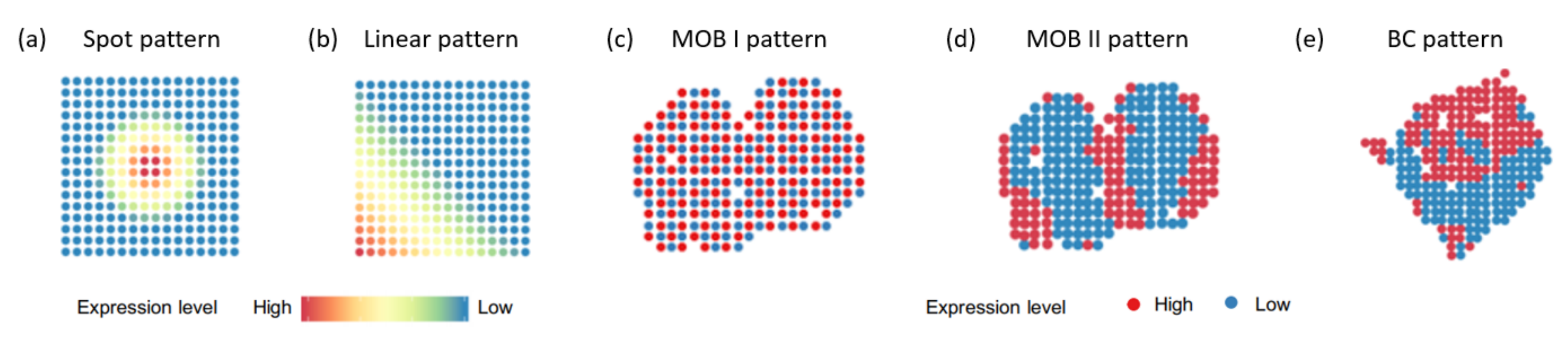}
	\caption{Simulation study: The five spatial patterns used to generate the simulated data. (a) and (b) Two artificial patterns with the Gaussian and linear kernel; (c) An artificial pattern with complete attraction pattern, i.e. $\theta\rightarrow\infty$; (e) and (d) Two real patterns constructed from the mouse olfactory bulb (MOB) and human breast cancer (BC) study.}
	\label{pattern}
\end{figure}

We generated simulated data from three artificial spatial patterns (see Figure \ref{pattern}(a)-(c)) and two real spatial patterns (see Figure \ref{pattern}(d) and (e)). The two real patterns were constructed from the mouse olfactory bulb (MOB) and human breast cancer (BC) datasets analyzed in this paper. The first two artificial patterns named spot and linear were on a $16\times16$ square lattice ($n=256$ spots), while the remaining one named MOB I was on $n=260$ spots. The MOB II and BC patterns were on $n=260$ and $250$ spots, respectively. We set $p=100$, among which $15$ were SV genes. We followed the data generative schemes \citep{edsgard2018identification,sun2020statistical, li2020bayesian} to simulate the gene expression count table $\boldsymbol{Y}$, which was substantially different from the model assumptions of BOOST-MI. For each gene $j$, the log relative expression level at spot $i$ was generated via \[\log\tilde{y}_{ij}=\begin{cases}\begin{array}{rl}\beta_0+e_i+\epsilon_{ij} & \quad\text{if gene } j \text{ is an SV gene}\\ e_i+\epsilon_{ij} & \quad \text{otherwise}\end{array}\end{cases},\] where $\beta_0$ denotes the baseline relative expression level and $\epsilon_{ij}$ denotes the non-spatial errors following $\text{N}(0,\sigma_\epsilon^2)$. We set $\beta_0=2$ and $\sigma_\epsilon=0.3$. For a non-SV gene, the relative expression levels were from a log-normal (LN) distribution with mean and variance being $2$ and $0.3^2$. Consequently, no spatial correlation should be observed. For an SV gene with the spot pattern, the values of $e_i$’s of the four center spots at $(8,8)$, $(8,9)$, $(9,8)$, and $(9,9)$ were set to $\log6$, while all others were linearly decreased to zero within a radius of five spots. For an SV gene with the linear pattern, the value of $e_i$ of the most bottom-left spot at $(1,1)$ was set to $\log6$, while all others were linearly decreased to zero along the diagonal line. For an SV gene with the remaining patterns, each spot was dichotomized into low and high-expression levels with $e_i=0$ and $\log3$, respectively. To mimic the excess zeros and over-dispersion in the real ST datasets, we sampled each gene expression count $y_{ij}$ from a zero-inflated negative binomial (ZINB) model, $y_{ij}\sim\pi_iI(y_{ij}=0)+(1-\pi_i)\text{NB}(s_i\tilde{y}_{ij},\phi_j)$, where the size factor $s_i\sim \text{LN}(0,0.2^2)$ and the dispersion parameter $\phi_j$ was from an exponential distribution with mean $10$. For the choice of the false zero proportion $\pi_i$, we randomly selected $10\%$, $30\%$, or $50\%$ counts and forced their values to zero. Combined with the five patterns and three zero-inflation settings, there were $15$ different scenarios. For each scenario, we repeated the above steps to generate $10$ replicates.

We chose to normalize the raw counts $\boldsymbol{Y}$ using TSS and dichotomize the relative expression levels for each gene using GMM as the default setting. As for BOOST-MI, the prior specification are $\omega_0\sim\text{N}(1,\sigma_{\omega}^2)$ and $\theta\sim\text{N}(0,\sigma_\theta^2)$. We set $\sigma_{\omega}=2.5$ and $\sigma_\theta=1/3$. The former indicated that the underlying proportion of the low-expression spots was expected between $1\%$ and $99\%$ with a probability of $95\%$, while the latter ensured about $99\%$ of $\theta_j$'s value ranged from $-1$ to $1$ a priori. A follow-up sensitivity analysis indicated that BOOST-MI was incredibly insensitive to the choice of these two hyperparameters (see Section S3 and Figure S4 in the supplement). As for the MCMC algorithm, we ran four independent MCMC chains for each gene with $10,000$ iterations, discarding the first half as burn-in. We started each chain from a model by randomly drawing all parameters from their prior distributions. Results we report below were obtained by pooling together the MCMC outputs from the four chains.

Both BOOST-MI and BOOST-GP identify SV genes based on BFs, while BinSpect, SPARK and SpatialDE output $p$-values to guide the selection. First, we used the area under the curve (AUC) of the receiver operating characteristic (ROC) to evaluate the performance of all methods. The ROC curve was created by plotting the true positive rate against the false positive rate across various thresholds used to select SV genes based on BFs or $p$-values. Second, we classified each gene as an SV or non-SV gene by pinpointing a specified threshold. Specifically, we set the BF cutoff to $150$, corresponding to a decisive strength of evidence. To control the type-I error rate, we adjusted $p$-values from SPARK, SpatialDE, and BinSpect using the Benjamini-Hochberg method \citep{benjamini1995controlling} and chose a significance level of $0.05$ as the cutoff. We chose the Matthews correction coefficient (MCC) \citep{matthews1975comparison} as the secondary performance metric, because SV genes is usually only a small subset of all genes, making other binary classification metrics not suitable. AUC yields a value between $0$ and $1$, and MCC value ranges from $-1$ to $1$. For both of them, the larger the value, the more accurate the identification.

\begin{figure}[!p]
	\centering
	\includegraphics[width = \textwidth]{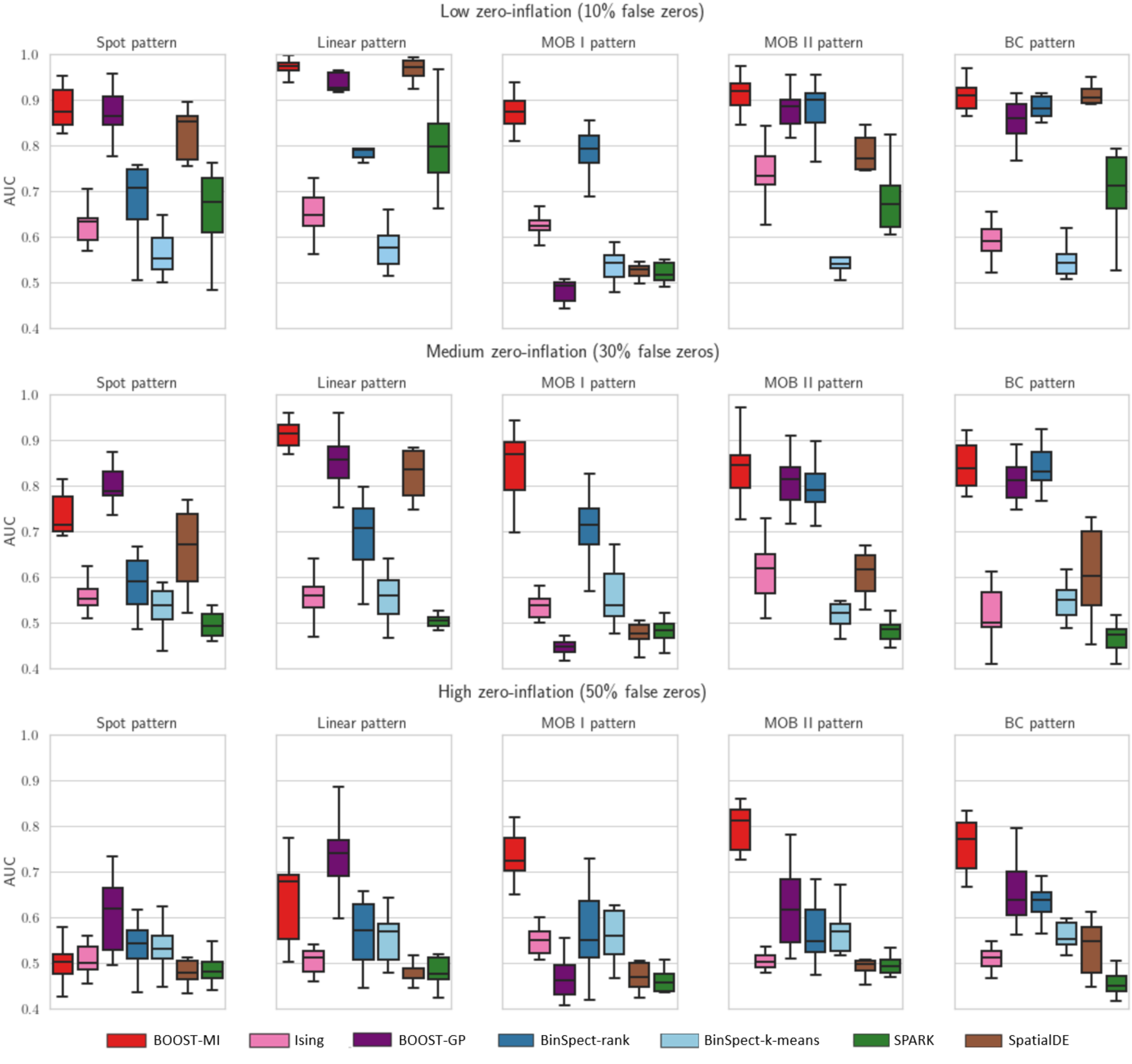}
	\caption{Simulation study: The boxplots of AUCs achieved by BOOST-MI, Ising, BOOST-GP, BinSpect, SPARK, and SpatialDE under different scenarios in terms of spatial pattern and zero-inflation setting.}
	\label{simu_auc}
\end{figure}

According to Figure \ref{simu_auc}, which displays the boxplots of AUCs by different methods over ten replicated datasets under each scenario, we concluded as follows. First, BOOST-MI achieved the highest performance in terms of median AUC under $12$ out of $15$ scenarios, while BOOST-GP only had a marginal advantage over BOOST-MI under the scenario with a medium or high proportion of false zeros and the artificial spot or linear pattern. Second, under the low zero-inflation setting, SPARK had a similar performance with BOOST-MI when the SV genes were generated from linear and BC patterns. However, it suffered from reduced power under the medium and high zero-inflation settings. This clearly suggested that realistic modeling in BOOST-MI contributed to its advantage over all other methods. Third, BinSpect was very sensitive to the choice of clustering methods. For example, the one based on top percentage rank performed significantly better than the one based on $k$-means under almost all scenarios, excluding the four scenarios with a high proportion of false zeros and the first four spatial patterns. In contrast, BOOST-MI is considerably robust to different normalization and dichotomization methods (see Figure S1 and S2 in the supplement). Forth, our BOOST-MI with the modified Hamiltonian consistently outperformed the Ising model with the classic Hamiltonian. Last but not least, SPARK, SpatialDE and BOOST-GP absolutely had no power to detect SV genes with the MOB I pattern, which was defined by a large positive interaction parameter in the Ising model. They might miss some important discoveries in real data analysis. Meanwhile, BinSpect had no satisfactory performance to detect such a pattern either, indicating our model-based analysis BOOST-MI could sharpen inferences. Besides, we reported the summaries of all methods' performance in terms of MCCs in Figure S5 and Table S3 in the supplement, respectively. Those results led to similar conclusions.

Regarding the efficiency, the average execution time per gene was $0.004$, $0.290$, and $0.067$ seconds for BinSpect, SPARK, and SpatialDE. BOOST-MI spent $4.5$ seconds per gene on average due to the computationally intensive DMH algorithms. In contrast, BOOST-GP had around two times higher computational cost than BOOST-MI, which spent $11.3$ seconds per gene on average. All experiments were implemented on a high performance computing server with two Intel Xeon CPUs ($45$ MB cache and $2.10$ GHz) and $250$ GB memory.

\subsection{Application to mouse olfactory bulb ST dataset} \label{mouse_section}

The first dataset that we applied BOOST-MI to analyze is a publicly available ST dataset in a mouse olfactory bulb (MOB) study. It is accessible on the Spatial Research Lab (\url{http://www.spatialresearch.org}). There are $12$ replicates in this study. Following the previous studies \citep{svensson2018spatialde,sun2020statistical,li2020bayesian}, we used the MOB replicate 11, which contains $16,218$ genes measured on $262$ spots. We applied the quality control steps described below before applying BOOST-MI. First, we excluded spots with fewer than ten total counts across all genes. Then, those genes with more than $80\%$ zero read counts across all spots were dropped. After these two steps, the MOB data had $n=260$ spots and $p=9,769$ genes. For other methods, we applied the pre-processing procedures suggested in their papers. We used the same prior specification, algorithm setting, and significance criteria used in the simulation study. We ran four independent MCMC chains and used the potential scale reduction factor (PSRF) \citep{gelman1992inference} to diagnose MCMC convergence. PSRF is a statistic comparing the estimated between-chains and within-chain variances for a model parameter. Its value should be close to one if multiple chains have converged to the target posterior distribution. The PSRFs for all $\omega_{j0}$'s and $\theta_j$'s were below $1.1$, clearly suggesting that the MCMC algorithms converged. Then, for each dataset, we pooled together the outputs from the four chains and selected SV genes based on BFs. We only compared BOOST-MI with SPARK and BinSpect-rank due to the poor performance of SpatialDE and heavy computational burden of BOOST-GP.

\begin{figure}[!p]
    \centering
    \includegraphics[width = 0.9\textwidth]{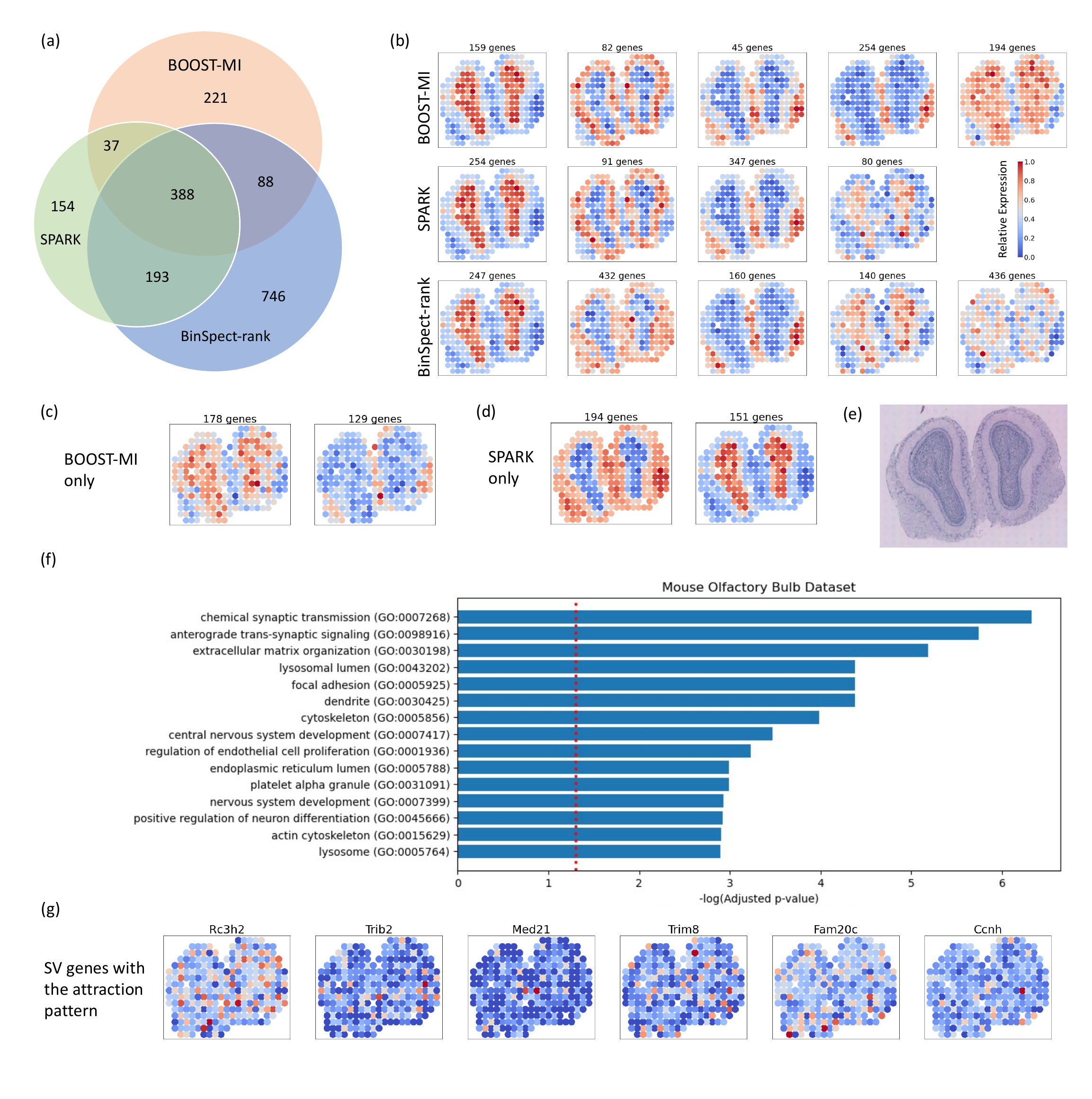}
    \caption{ \footnotesize Real data analysis on the mouse olfactory bulb (MOB) dataset: (a) The Venn diagram of SV genes identified by BOOST-MI, SPARK, and BinSpect-rank; (b) Distinct spatial expression patterns summarized on the basis of $734$,  $772$, and $1,415$ SV genes identified by BOOST-MI, SPARK, and BinSpect-rank; (c) Distinct spatial expression patterns summarized on the basis of $307$ SV genes identified by BOOST-MI only; (d) Distinct spatial expression patterns summarized on the basis of $345$ SV genes identified by SPARK only; (e) The associated hematoxylin and eosin (H\&E)-stained tissue slides of the analyzed MOB dataset; (f) Top 15 terms of gene ontology (GO) enrichment analysis of $734$ SV genes identified by BOOST-MI in MOB data, with red dashed line indicating a significance level of $0.05$; (g) The top six genes with the attraction pattern in terms of BF values.}
    \label{mob_result}
\end{figure}

BOOST-MI identified $734$ SV genes, which was approximately the same number of SV genes detected by SPARK ($772$ SV genes) and around half of the number of SV genes detected by BinSpect ($1,415$ SV genes). Figure \ref{mob_result}(a) is a Venn diagram showing the overlap of detected SV genes by all methods. More than half of the SV genes identified by BOOST-MI ($388$ out of $734$) were also reported by the other two, while there were $221$ unique SV genes to BOOST-MI. Most of the SV genes detected by SPARK ($618$ out of $772$) were also included in the result of BOOST-MI or BinSpect, which indicates that SPARK is a relatively conservative method for SV gene identification. BinSpect was shown to be the most aggressive method that reported the most SV genes, more than half of which ($746$ out of $1,415$) were not founded by either alternative.

To further explore the spatial patterns we had found, we performed the agglomerative hierarchical clustering on the SV genes identified by each of the three methods. For better visualization, we followed the data preprocessing \citep{svensson2018spatialde,sun2020statistical} to normalize the raw read counts to relative expression levels using log-VST. Next, based on the distance matrix computed from relative expression levels of all pairs of SV genes, a hierarchical clustering dendrogram was constructed. We then determined the number of clusters by cutting the hierarchical clustering dendrogram at a height corresponding to a clear separation. Last, we summarized the expression patterns via the averaged relative expression levels within each cluster. As a result, the SV genes detected by BOOST-MI and BinSpect were clustered into five groups, while there were four groups for SPARK, shown in Figure \ref{mob_result}(b). Consistent with results in a previous study \citep{sun2020statistical}, three major spatial patterns were shown in the first three columns in Figure \ref{mob_result}(b). Note that the fourth pattern of SPARK (with $80$ genes) and BinSpect (with $140$ genes) could be merged into their first pattern, respectively. Also, the third and fourth patterns of BOOST-MI had a high similarity. It is noteworthy that a unique pattern (the last pattern of BOOST-MI) could only be detected by BOOST-MI with $194$ SV genes, indicating our method had a higher power. To further compare BOOST-MI with SPARK, we repeated the above procedure on those SV genes detected by BOOST-MI or SPARK only. The results are shown in Figure \ref{mob_result}(c) and (d), respectively. The $307$ genes identified only by BOOST-MI could be categorized into two groups, including the unique pattern with $178$ genes. SV genes only detected by SPARK showed a strong periodic pattern, suggesting that SPARK might be more sensitive to the smooth periodic spatial pattern. 

Next, using a Python wrapper GSEAPY \citep{subramanian2007gsea,kuleshov2016enrichr}, we performed gene ontology (GO) enrichment analysis of the $734$ SV genes identified by BOOST-MI to explore their relevant biological functions. A total of $3,929$ mouse GO terms in three components (biological processes, cellular components, and molecular functions) had at least one gene overlap. Controlling the false discovery rate (FDR) at $5\%$, we found $155$ GO terms, the top $15$ of which (with the smallest $p$-values) are shown in Figure \ref{mob_result}(f). As with SPARK, many enriched gene sets were related to synaptic signaling and the nervous system, both of which are significantly associated with the synaptic organization and olfactory bulb development \citep{treloar2010development}. Examples include chemical synaptic transmission (GO:0007268; adjusted $p$-value $4.69\times 10^{-7}$) and nervous system development (GO:0007399; adjusted $p$-value $1.18\times 10^{-3}$).

Last but not least, BOOST-MI identified $60$ genes that had an attraction pattern with a positive interaction parameter in the Ising model. Table S4 in the supplement lists all of them. To analyze the potential biological functions of these SV genes, we performed functional enrichment analysis. There were $603$ mouse GO terms and $43$ Kyoto Encyclopedia of Genes and Genomes (KEGG) terms with at least one gene overlapping with those SV genes. We also found some statistically significant GO and KEGG terms, which had adjusted $p$-values less than a significance level of $0.05$. For example, holo TFIIH complex (GO:0005675) and nucleotide excision repair KEGG term were significantly enriched (adjusted $p$-value $0.040$ and $0.013$, respectively). In all, these discoveries highlighted the advantage of BOOST-MI.

\subsection{Application to human breast cancer ST dataset}\label{human}

The second dataset is an available ST dataset in a human breast cancer (BC) study, which is also accessible on the Spatial Research Lab (\url{http://www.spatialresearch.org}). There are four layers available and we used the BC layer 2, which contains $14,789$ genes measured on $251$ spots. We applied the same quality control steps described in Section \ref{mouse_section}. The BC data had $n=250$ spots and $p=2,280$ genes. We applied the same pre-processing procedures for other methods and the same prior specification, algorithm setting, and significance criteria as mentioned. To check convergence, we run four chains and the PSRFs for all $\omega_{j0}$'s and $\theta_j$'s were below $1.1$. We also compared our result with BinSpect and SPARK.

\begin{figure}[!p]
    \centering
    \includegraphics[width = \textwidth]{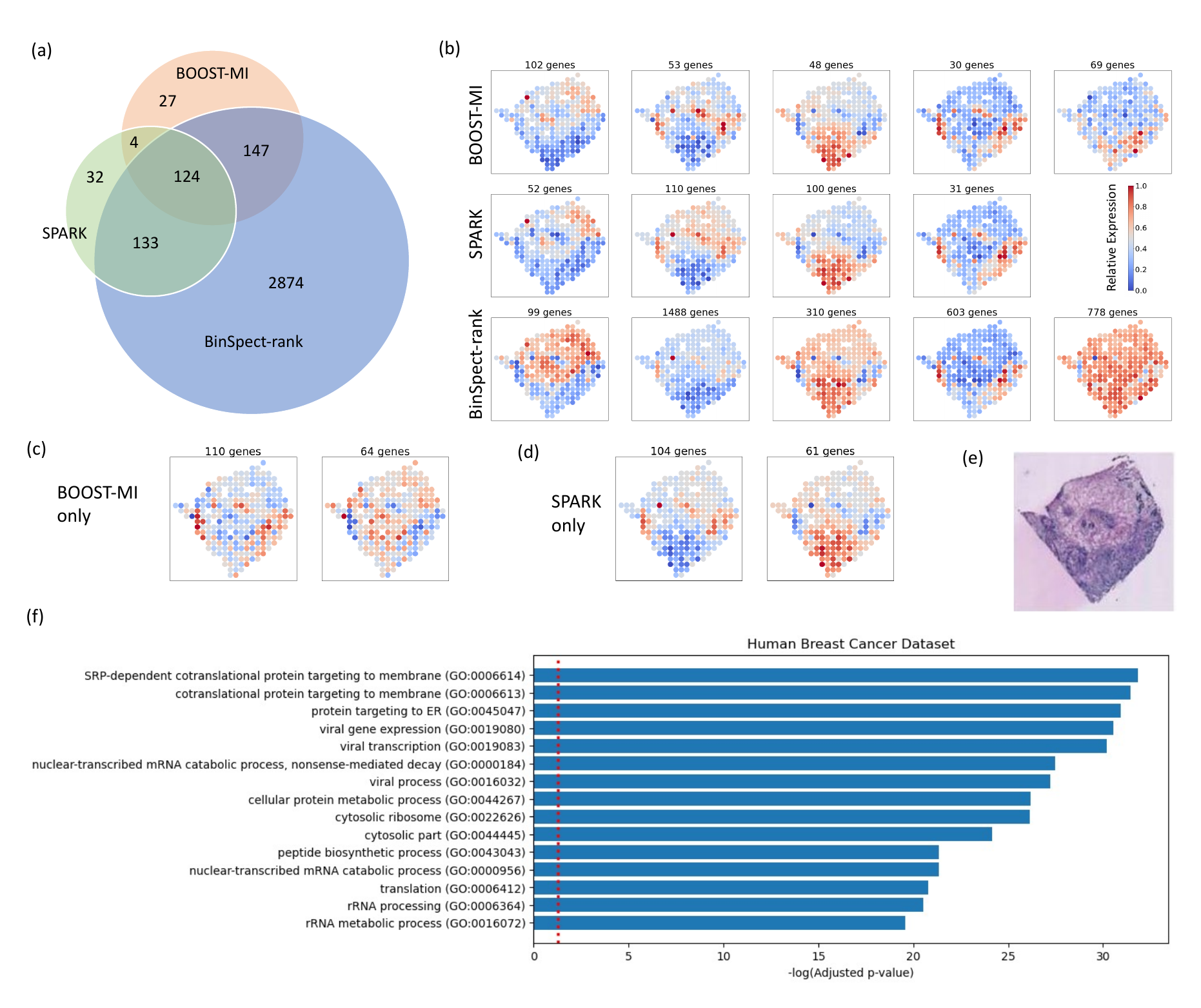}
    \caption{\footnotesize Real data analysis on the human breast cancer (BC) dataset: (a) The Venn diagram of SV genes identified by BOOST-MI, SPARK, and BinSpect-rank; (b) Distinct spatial expression patterns summarized on the basis of $302$,  $293$, and $3,278$ SV genes identified by BOOST-MI, SPARK, and BinSpect-rank; (c) Distinct spatial expression patterns summarized on the basis of $174$ SV genes identified by BOOST-MI only; (d) Distinct spatial expression patterns summarized on the basis of $165$ SV genes identified by SPARK only; (e) The associated hematoxylin and eosin (H\&E)-stained tissue slides of the analyzed BC dataset; (f) Top 15 terms of gene ontology (GO) enrichment analysis of $302$ SV genes identified by BOOST-MI in MOB data, with red dashed line indicating a significance level of $0.05$. }
    \label{bc_result}
\end{figure}

There were $302$ SV genes identified by BOOST-MI, which was slightly larger than the number of SV genes detected by SPARK ($292$ SV genes) and around one-tenth of the number of SV genes detected by BinSpect ($3,278$ SV genes). A Venn diagram, as shown in Figure \ref{bc_result}(a), indicates that over $40\%$  of the SV genes detected by BOOST-MI ($124$ out of $302$) were in common with the result obtained by both SPARK and BinSpect. There were only $27$ SV genes identified only by BOOST-MI. Once again, it demonstrated that SPARK is a conservative method and Binspect is an aggressive one, since nearly $90\%$ of the SV genes detected by SPARK ($260$ out of $293$) were also included in the results of BOOST-MI or BinSpect, while nearly $90\%$ of the SV genes reported by BinSpect ($2,874$ out of $3,278$) were found by neither BOOST-MI nor SPARK.

We reported more detailed analysis results, following the same procedure when studying the MOB dataset. Each row in Figure \ref{bc_result}(b) shows the distinct expression patterns detected by BOOST-MI, SPARK, and BinSpect, respectively. There were five, four, and five groups obtained by performing the agglomerative hierarchical clustering. It is noteworthy that BOOST-MI detected all four patterns discovered by SPARK. However, BOOST-MI detected $102$ SV genes with the first pattern, while SPARK reported only $52$ SV genes with a similar pattern. BOOST-MI identified fewer SV genes than SPARK for the second and third patterns, while the number of SV genes was almost the same for the fourth pattern. The last pattern of BOOST-MI was unique, and it was approximately the reversed pattern of the first one. Additionally, we repeated the same procedure on the SV genes only detected by BOOST-MI and SPARK, which is shown in Figure \ref{bc_result}(c) and (d). In this BC study, BOOST-MI reported no SV genes with an attraction pattern.

Finally, we performed GO enrichment analysis of the $302$ SV genes identified by BOOST-MI. A total of $2,477$ human GO terms had at least one gene overlap with those identified SV genes. At an FDR of $5\%$, $116$ GO terms were found. Figure \ref{bc_result}(f) shows the top $15$ GO terms with the smallest adjusted $p$-values. SPARK discovered many enriched gene sets, which were related to extracellular matrix organization and immune responses \citep{sun2020statistical}. Although these GO terms were not shown in Figure \ref{bc_result}(f), BOOST-MI did detect the same related terms with a significant result (e.g. the adjusted $p$-value for extracellular matrix organization (GO:0030198) was $1.53\times 10^{-14}$). Furthermore, more virus-related GO terms were found to be significant in our analysis. There is strong evidence that many types of the virus may have a causal relationship with human breast cancers \citep{lawson2010viruses}. For example, virus life cycle GO term (GO:0019058) was significantly enriched in the reported SV genes by BOOST-MI, while not statistically significant for SV genes identified by SPARK (the adjusted $p$-value was only $0.829$).

\section{Conclusion}
\label{sec4}

In this paper, we develop a multi-stage method to identify SV genes from ST data. Instead of characterizing an SV gene via a pre-specified kernel by most existing methods, we define a spatial pattern via the Hamiltonian energy in the Ising model with external fields. BOOST-MI offers the flexibility to choose different normalization and dichotomization methods, and it is considerably robust to various biological and technique noise and biases. In the simulation study, BOOST-MI had a noticeable advantage over alternative methods, especially when there were a great number of zeros in the data. This is very encouraging since BOOST-MI does not directly model excess zeros. BOOST-MI led to more discoveries in real data analysis, such as novel spatial patterns that had never been reported and novel SV genes that kernel-based methods are unable to detect. 

Several extensions of our model are worth investigating. First of all, the proposed model could be extended to model $k$ discrete gene expression levels via a Potts model or its modified version to characterize finer spatial patterns. The number of components $k$ can even be estimated \citep{green2002hidden}. Second, based on the multi-stage BOOST-MI, a full hierarchical Bayesian framework could be developed to directly model the ST count data to sharpen inference. For example, with a joint Bayesian inference, BOOST-MI could further incorporate pathway information as prior knowledge to integrate the regulatory relationships between genes to perform a joint selection on SV genes. Moreover, since our approach requires the gene expression measured on a lattice grid, it is necessary to generalize our model to detect SV genes from SMP data produced by other platforms based on the single-molecule fluorescence in situ hybridization (FISH), such as sequential FISH (seqFISH) \citep{lubeck2014single} and multiplexed error-robust FISH (MERFISH) \citep{chen2015spatially}, through the hidden Ising or Potts models \citep{li2019bayesian}. Those SMP techniques measure the expression levels of hundreds of genes on thousands of cells, which are irregularly scattered in a planar space. Finally, it is possible to investigate other approximate Bayesian computation methods to reduce the computational cost of BOOST-MI. These future directions could potentially further improve the performance of BOOST-MI.

\section*{Software}
\label{sec5}

Two publicly available ST datasets in a mouse olfactory bulb study and a human breast cancer study are accessible on the Spatial Research Lab (\url{http://www.spatialresearch.org}).
All simulated and real data used for analysis, and the related source code in \texttt{R}/\texttt{C++} are available at \url{https://github.com/Xijiang1997/BOOST-MI}.

\section*{Acknowledgments}

Computational support was generously provided by Southern Methodist University Center for Research Computing. The authors would like to thank Suhana Bedi from the University of Texas at Dallas for helping us in implementing the normalization methods, and Jessie Norris from the University of Texas Southwestern Medical Center for helping us in proofreading the manuscript. 

\section*{Funding}
This study was partially supported by the National Institutes of Health (NIH) [1R01GM140012, 1R01GM141519, P30CA142543] and the Cancer Prevention and Research Institute of Texas [RP190107]. The funders had no role in the design of the study and collection, analysis, and interpretation of data or in writing the manuscript. 

\newpage
\bibliographystyle{biorefs}
\bibliography{refs}


\newpage 

\section*{SUPPLEMENTARY NOTES}
\setcounter{figure}{0}

\renewcommand{\thetable}{S\arabic{table}}%

\renewcommand{\thesubsection}{S\arabic{subsection}}

\subsection{Full details of data normalization}

Normalization is critical to the analysis of sequence count data that suffer from various sequence artifacts and bias. We provide eight normalization methods in two categories.

The first type is based on size factor estimation. Let $s_i$ be the size factor of sample $i$, capturing all nuisance effects. Each relative gene expression level can be computed as $\tilde{y}_{ij} = y_{ij}/s_i$. If the main interest is in the absolute gene expression level, then all $s_i$'s are set to the same value (e.g. $s_1=\ldots=s_n=1$); otherwise, we compute $s_i$'s directly from the gene expression count data. The simplest way is to set $s_i\propto Y_i=\sum_{j=1}^py_{ij}$, i.e., the total number of counts across all genes in each sample (known as sequencing depth or library size). Note that SPARK \citep{sun2020statistical} suggests this normalization, namely total sum scaling (TSS). In practice, we could consider other estimations on $s_i$'s, which have been introduced for mitigating the influence of extremely low and high counts when analyzing bulk RNA-seq data, such as upper-quartiles (Q75) \citep{bullard2010evaluation}, relative log expression (RLE) \citep{anders2010differential}, and weighted trimmed mean by M-values (TMM) \citep{robinson2010scaling}. Table \ref{size factor} provides the definitions of the above size factor estimations. The size factor estimation is usually combined with some constraint, such as $\prod_{i=1}^ns_i=1$. 

The other type of normalization method is based on variance-stabilizing transformation (VST), which aims to transform a random variable with a negative binomial distribution into one with an approximately normal distribution. There are three options: Na\"ive, Anscombe \citep{anscombe1948transformation}, and logarithm, namely N-VST, A-VST, and log-VST, all of which can be abstracted as $\tilde{y}_{ij}=g(y_{ij},\phi)$, where $g$ is the transformation-specific function (see Table \ref{vst}) and $\phi$ is the dispersion parameter estimated from the count data. Then, the relative gene expression levels are further adjusted for the log-scale total read counts, i.e. $\log Y_i$, via a linear regression model. Note that SpatialDE \citep{svensson2018spatialde} employs the log-VST normalization before fitting the geostatistical model. 

We conducted a sensitivity analysis to investigate how different normalization methods affect the SV gene identification. We simulated ten replicated datasets following the data generating process described in Section 5.1 in the manuscript. We include only the scenario with the medium zero-inflation setting ($30\%$ false zeros) and the two real spatial patterns (MOB II and BC). We assessed the model performance in terms of the area under the curve (AUC). The result is summarized in Figure \ref{sensi_norm}. We found that the methods based on size factor estimation significantly outperformed the VST-based methods. BOOST-MI was robust to the four size factor-based normalization methods. We conducted the analysis of variance (ANOVA) test on all pairs of the four size factor-based normalization methods. All $p$-values were greater than $0.05$, confirming no significant difference among the four choices.

\subsection{Full details of data dichotomization}
After correcting for sequence artifacts and bias, we denoise the relative expression levels by partitioning all spots into two groups. This step outputs the suitable data type required in the subsequent analysis and makes BOOST-MI more robust in the face of over-dispersion and zero-inflation.

There is no consensus on the dichotomization of spots based on either absolute or relative expression level. BinSpect \citep{dries2021giotto} suggests allocating those spots with the top $30\%$ relative expression levels to the high-expression group and the remaining to the low-expression group. Meanwhile, it also considers $k$-means ($k=2$) as an alternative to avoid choosing a hard percentage rank cutoff. 

We provide two choices of clustering methods. The first one is similar in spirit to BinSpect-$k$-means, searching the $\boldsymbol{p}_j$ corresponding to the minimum within-cluster sum of squares,
\[\arg\min_{\boldsymbol{p}_j}\sum_{k}\sum_{i}I(p_{ij}=k)(\tilde{y}_{ij}-m_k)^2,\]
where $I(\cdot)$ denotes the indicator function, $m_k=\sum_{i}\tilde{y}_{ij}I(p_{ij}=k)/n_k$ and $n_k=\sum_{i}I(p_{ij}=k)$ are the sample mean and size of each group. However, before applying $k$-means, we first exclude those spots whose $\tilde{y}_{ij}$'s are larger than $\tilde{q}_j^{0.75}+3\left(\tilde{q}_j^{0.75}-\tilde{q}_j^{0.25}\right)$, where $\tilde{q}_j^x$ is defined as the $x$-th sample quantile of the relative expression levels of gene $j$, and then removes those spots with $y_{ij}=0$. Note that in the context of box-and-whisker plotting, a data point is defined as an extreme outlier if it stands outside this limit. We directly allocate those discarded spots in the first and second steps to the low and high-expression groups, respectively. Compared with BinSpect, this additional preprocessing leads to a more robust performance when excessive zeros and outliers are presented. 

In addition to $k$-means that is implicitly based on pairwise distances between relative expression levels, we propose to estimate $\boldsymbol{p}_j$ via fitting a two-component Gaussian mixture model (GMM) with unequal variances, 
\[\tilde{y}_{ij}|p_{ij},\boldsymbol{\mu}_j,\boldsymbol{\sigma}_j^2\sim (1-p_{ij})\text{N}(\mu_{j0},\sigma_{j0}^2)+p_{ij}\text{N}(\mu_{j1},\sigma_{j1}^2),\]
subjecting to $\mu_{j0}<\mu_{j1}$. Here $\boldsymbol{\mu}_j=(\mu_{j0},\mu_{j1})$ and $\boldsymbol{\sigma}_j^2=(\sigma_{j0}^2,\sigma_{j1}^2)$ are the group means and variances that need to be estimated. To ensure the dichotomized expression levels are of the best quality to perform the subsequent modeling, we implement the above filtering steps as well.

To evaluate the performance between the two choices, we conducted a sensitivity analysis. We simulated ten replicated datasets following the data generating process described in Section 5.1 in the manuscript. We include only the scenario with the medium zero-inflation setting ($30\%$ false zeros) and the two real spatial patterns (MOB II and BC). We assessed the model performance in terms of AUC. The result is summarized in Figure \ref{sensi_cluster}. BOOST-MI was robust to the two Dichotomization methods. We conducted the pairwise $t$-test. The resulting $p$-value$=0.585$ and $0.850$ for MOB II and BC pattern, respectively, confirming no significant difference among the two choices.

\subsection{Sensitivity analysis for the choices of BOOST-MI hyperparameters}

We conducted a sensitivity analysis to investigate the sensitivity of BOOST-MI to the choice
of $\sigma_\omega$ and $\sigma_\theta$. We applied BOOST-MI to each of the ten replicated datasets under the scenario with the low zero-inflation setting ($10\%$ false zeros) and MOB II pattern. We varied values of $\sigma_{\theta}$ from $1/4$ to $2$ and $\sigma_{\omega}$ from $1$ to $100$. We chose five values for each hyperparameter, resulting in $25$ combinations. We assessed the model performance in terms of both AUC and MCC, where the latter was based on a Bayes factor (BF) threshold of $150$. The result is summarized in Figure \ref{sensi_hyper}, clearly indicating that BOOST-MI was not sensitive to the choices of hyperparameters.

\renewcommand{\thefigure}{S\arabic{figure}}

\captionsetup[figure]{labelformat={default},name={FIGURE}, labelsep=colon}

\newpage
\begin{figure}[h]
	\centering
	\includegraphics[width = 14cm]{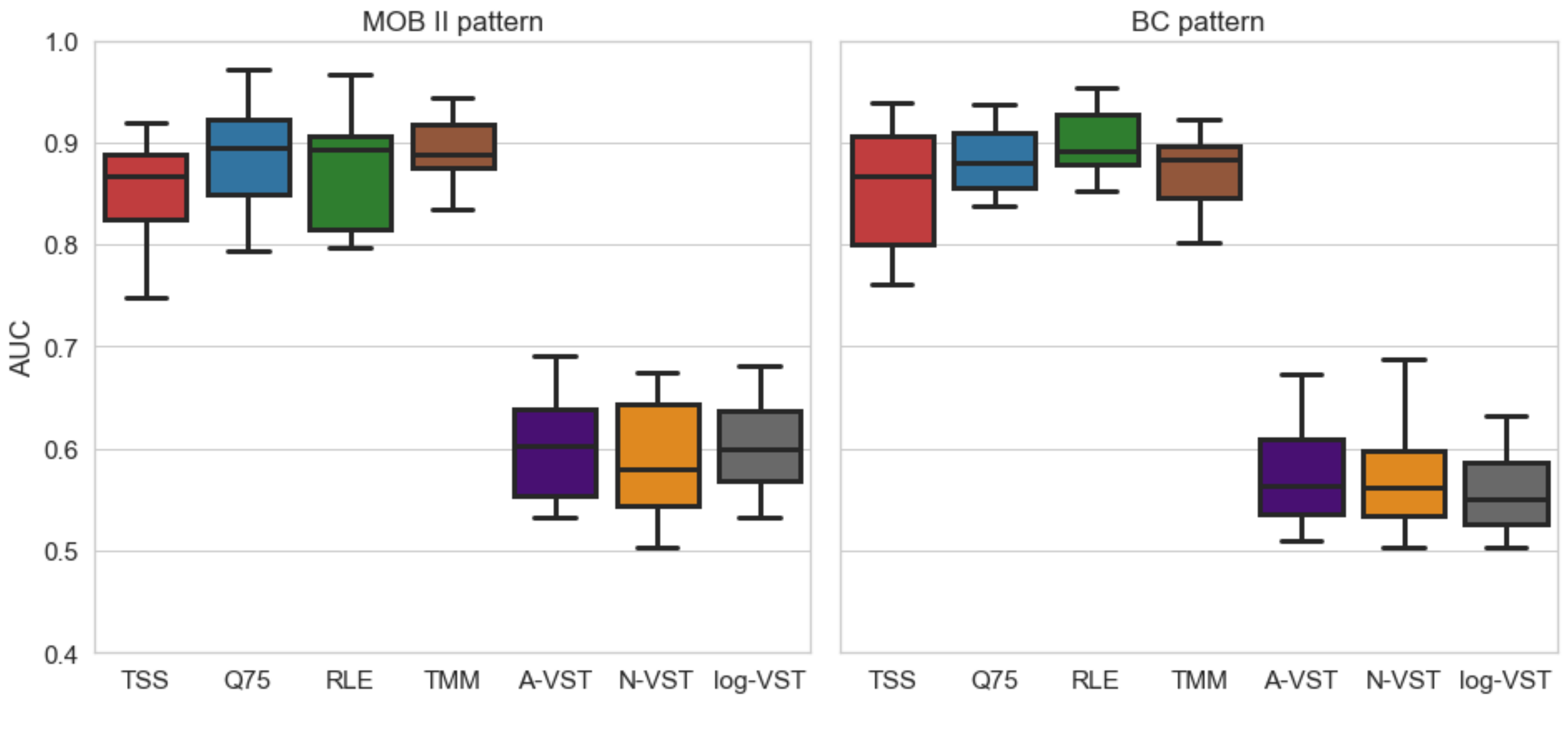}
	\caption{Simulation study: The boxplots of AUCs achieved by different normalization methods in BOOST-MI.}
	\label{sensi_norm}
\end{figure}

\begin{figure}[h]
	\centering
	\includegraphics[width = 10cm]{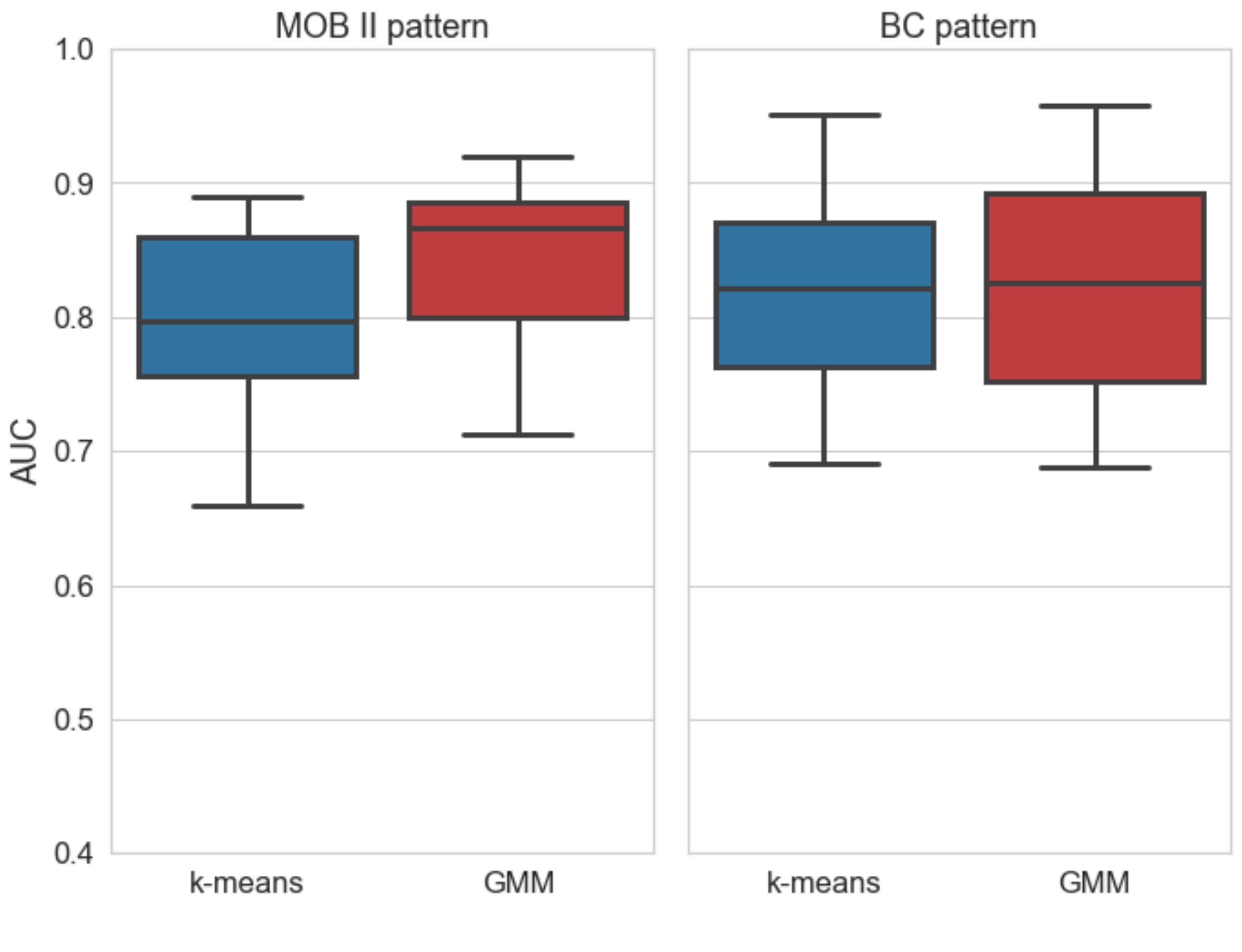}
	\caption{Simulation study: The boxplots of AUCs achieved by different dichotomization methods in BOOST-MI.}
	\label{sensi_cluster}
\end{figure}

\begin{figure}[h]
	\centering
	\includegraphics[width = 10cm]{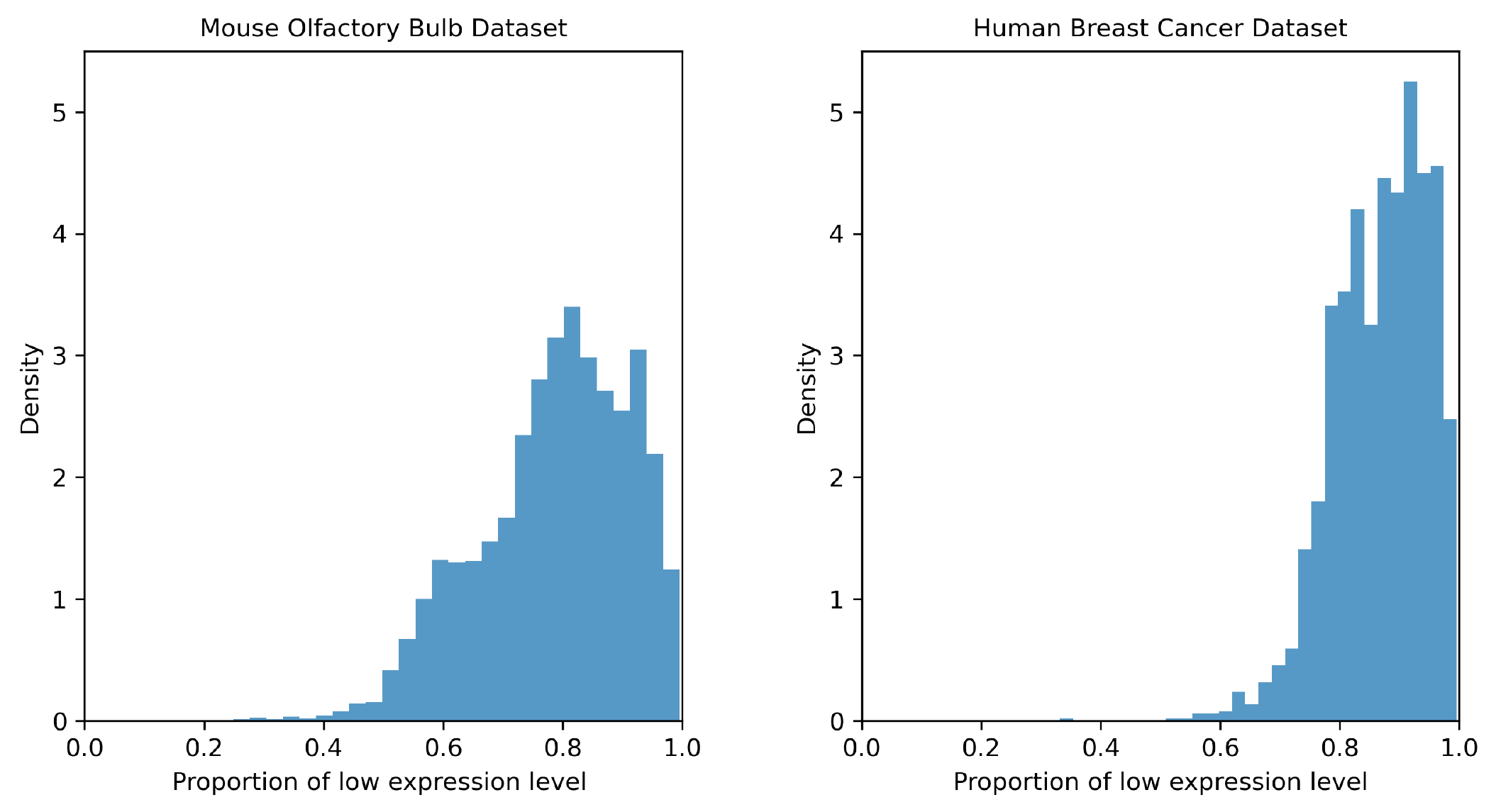}
	\caption{The histogram of low-expression spot proportions over all genes in the mouse olfactory bulb (MOB) and human breast cancer (BC) datasests.}
	\label{low}
\end{figure}

\begin{figure}[h]
	\centering
	\includegraphics[width = 15cm]{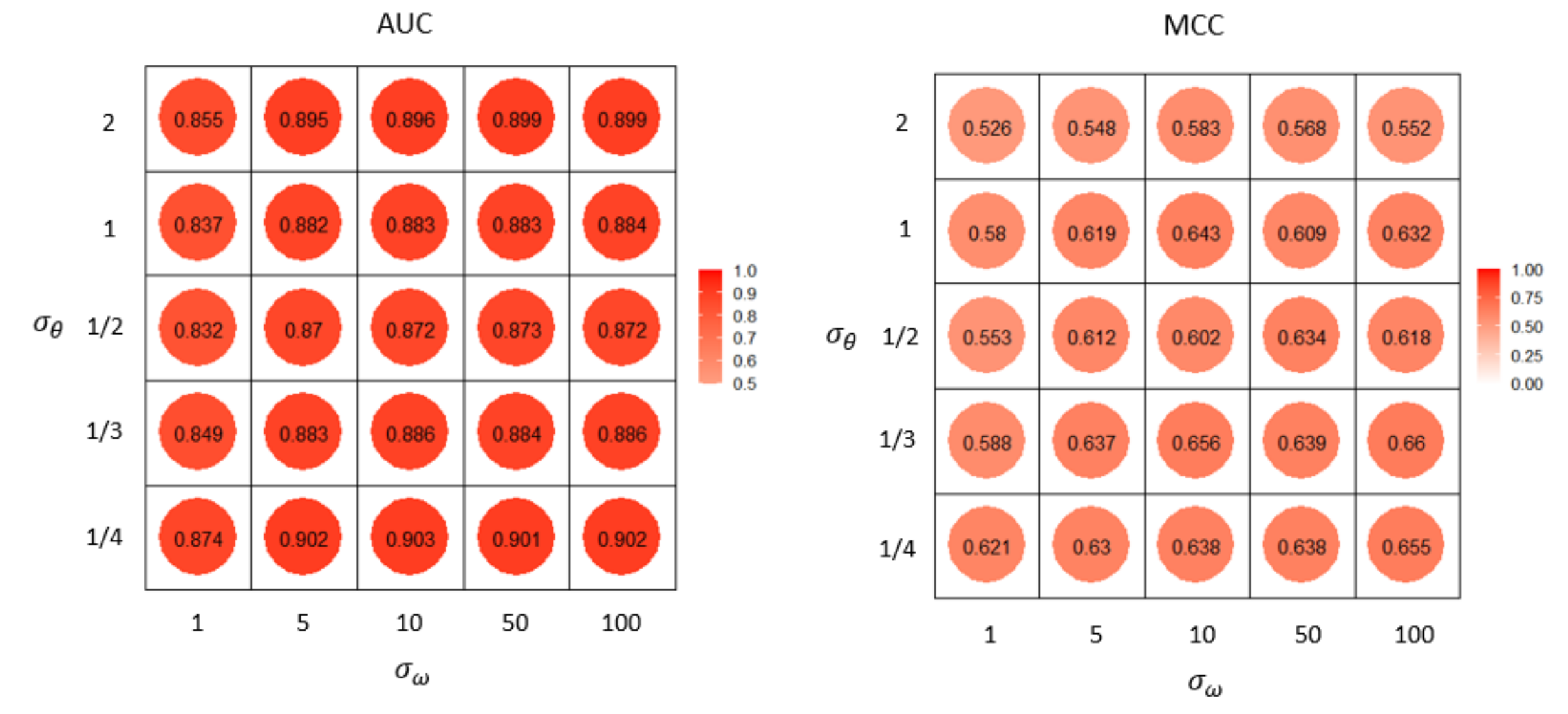}
	\caption{Simulation study: The heatmaps of averaged AUCs and MCCs achieved by different BOOST-MI hyperparameters $\sigma_\omega$ and $\sigma_\theta$.}
	\label{sensi_hyper}
\end{figure}

\begin{figure}[h]
	\centering
	\includegraphics[width = \textwidth]{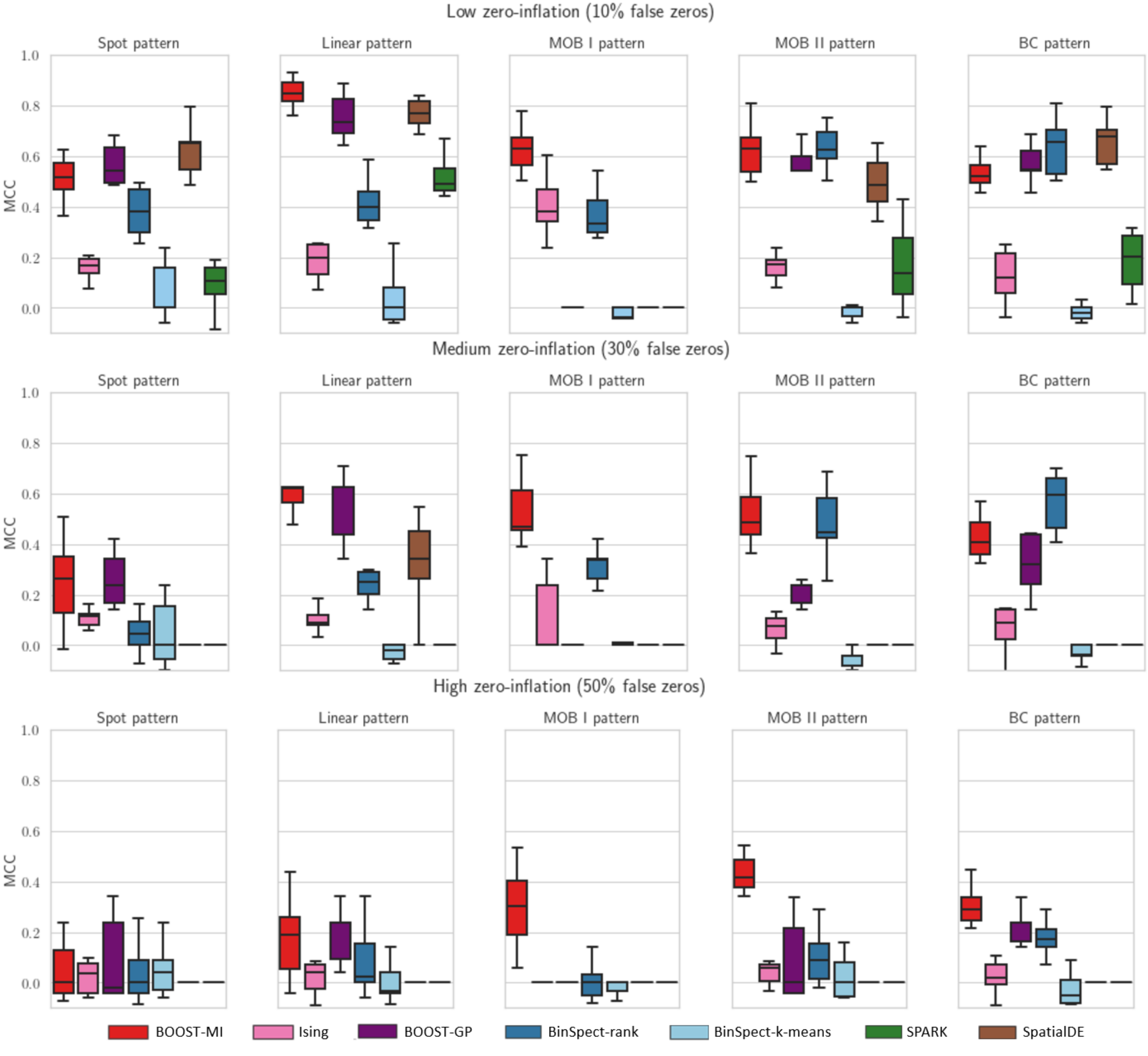}
	\caption{Simulation study: The boxplots of MCCs achieved by BOOST-MI, Ising, BOOST-GP, BinSpect, SPARK, and SpatialDE under different scenarios in terms of spatial pattern and zero-inflation setting.}
	\label{simu_mcc}
\end{figure}

\begin{table}[!h]
	\centering 
	\caption{List of size factors used for normalizing sequence count data.}\label{size factor}
	\renewcommand{\arraystretch}{1.2}
	\begin{tabular}{@{}l|l@{}}
		\hline
		Abbreviation & Definition \\\hline
		none & ${s}_i\propto 1$ \\\hline
		TSS  & ${s}_i\propto Y_{i\cdot}$ \\\hline
		Q75 \citep{bullard2010evaluation}  & ${s}_i\propto q_i^{0.75},$  \\\hline
		RLE \citep{anders2010differential}& ${s}_i\propto\text{median}_j\left\{y_{ij}/\sqrt[n]{\prod_{i'=1}^ny_{i'j}}\right\}$ \\\hline
		TMM \citep{robinson2010scaling} &
		${s}_i\propto Y_i\cdot\exp\left( \frac{\sum_{j\in G^*}\psi_j(i,r)M_j(i,r)}{\sum_{j\in G^*}\psi_j(i,r)}
		\right)$ \\\hline
		\multicolumn{2}{p{0.8\textwidth}}{\footnotesize Note 1: $q_i^x$ is defined as the $x$-th sample quantile of all the counts in sample $i$, i.e. there are $xp$ features in sample $i$ whose $y_{ij}$'s are less than $q_i^x$.}\\
		\multicolumn{2}{p{0.8\textwidth}}{\footnotesize Note 2: The $M$-value $M_j(i,r)=\log(y_{ij}/Y_{i})/\log(y_{rj}/Y_{r})$ and $A$-value $A_j(i,r)=(\log{y_{ij}/Y_{i}+\log{y_{rj}/Y_{r}}})/2$ are the ratio and average of log-scaled counts between sample $i$ and the reference sample $r$, respectively. $G^*$ denote a subset of genes whose $M$-values are not within the upper and lower $30\%$ of all $M$-values and $A$-values are not within the upper and lower $5\%$ of all $A$-values. The weight $\psi_{j}(i,r)=\frac{Y_{i}-y_{ij}}{y_{ij}Y_{i}}+\frac{Y_{r}-y_{rj}}{y_{rj}Y_{r}}$ is the inverse of the approximate asymptotic variances.}
	\end{tabular}
\end{table}

\begin{table}[!h]
	\centering
	\caption{List of variance-stabilizing transformations (VST) used for normalizing sequence count data.}\label{vst}
	\renewcommand{\arraystretch}{1.2}
	\begin{tabular}{@{}l|l@{}}
		\hline
		Abbreviation & Definition \\\hline
		A-VST  \citep{anscombe1948transformation}& $g(y_{ij},\phi)=\sinh^{-1}\sqrt{{y_{ij}}/{\phi}}$ \\\hline
		N-VST \citep{love2014moderated}& $g(y_{ij},\phi)=\sinh^{-1}\sqrt{\frac{y_{ij}+{3}/{8}}{{\phi}-{3}/{4}}}$ \\\hline
		log-VST \citep{anscombe1948transformation, svensson2018spatialde}& $g(y_{ij},\phi)=\log\left(y_{ij}+\phi/2\right)$ \\\hline
		\multicolumn{2}{p{0.8\textwidth}}{\footnotesize Note 1: $\sinh^{-1}(y)=\log(y+\sqrt{1+y^2})$.}\\
		\multicolumn{2}{p{0.8\textwidth}}{\footnotesize Note 2: $\phi$ can be estimated via a non-linear model: $s_j^2=\bar{y}_j+\bar{y}_j/\phi,j=1,\ldots,p$, where $\bar{y}_j$ and $s_j^2$  denote the sample mean and variance of all the counts belonging to gene $j$, i.e. $(y_{1j},\ldots,y_{nj})$.}
	\end{tabular}
\end{table}

\begin{table}[!h]
	\centering 
	\caption{Simulation study: The averaged MCCs (standard edivations) achieved by BOOST-MI, BOOST-GP, BinSpect, SPARK, and SpatialDE under different scenarios in terms of spatial pattern and zero-inflation setting.}
	\setlength{\tabcolsep}{6pt} 
	\renewcommand{\arraystretch}{0.7} 
	\begin{tabular}{l c c c c c}
		\hline
		\multirow{2}{*}{$\quad$} & \multicolumn{5}{c}{Low zero-inflation (10\% false zeros)} \\\cline{2-6}
		
		& Spot  & Linear  & MOB I    & MOB II  & BC\\\hline
		BOOST-MI & 0.519(0.071) & \textbf{0.853}(0.054) & \textbf{0.642}(0.103) & 0.626(0.086) & 0.531(0.108)  \\
		BOOST-GP & 0.551(0.118) & 0.726 (0.124) & 0.015(0.078) & 0.571(0.102) & 0.586(0.104) \\
		BinSpect-rank & 0.341(0.176) & 0.418(0.167)  & 0.333(0.172)& \textbf{0.632}(0.080) & 0.639(0.111) \\
		BinSpect-km & 0.061(0.117) & 0.036(0.102)& 0.041(0.158) & -0.019(0.035) & -0.021(0.032)  \\
		SPARK & \textbf{0.624}(0.091) & 0.768(0.053) & 0.000(0.000)& 0.488(0.116)& \textbf{0.652}(0.083)  \\ 
		SpatialDE   & 0.128(0.136) & 0.497(0.102) & 0.000(0.000)& 0.184(0.203)  & 0.276(0.197)\\
		\hline
		\multirow{2}{*}{$\quad$} & \multicolumn{5}{c}{Medium zero-inflation (30\% false zeros)} \\\cline{2-6}
		
		& Spot  & Linear  & MOB I  & MOB II  & BC  \\\hline
		BOOST-MI & 0.237(0.155) & \textbf{0.604}(0.077)& \textbf{0.568}(0.095) & \textbf{0.501}(0.157) & 0.398(0.071)  \\
		BOOST-GP & \textbf{0.254}(0.133) & 0.487 (0.115) & 0.000(0.000) & 0.239(0.117) & 0.387(0.136) \\
		BinSpect-rank & 0.067(0.111) & 0.242(0.055)& 0.306(0.096) & 0.478(0.139) & \textbf{0.567}(0.112)  \\
		BinSpect-km & 0.049(0.126) & -0.004(0.076) & 0.010(0.065) & -0.027(0.098) & -0.031(0.030) \\
		SPARK & 0.055(0.048) & 0.342(0.164)& 0.000(0.000) & 0.065(0.113)& 0.058(0.124)  \\ 
		SpatialDE   & 0.000(0.000) & 0.000(0.000) & 0.000(0.000)  & 0.000(0.000)& 0.000(0.000)\\
		\hline
		\multirow{2}{*}{$\quad$} & \multicolumn{5}{c}{High zero-inflation (50\% false zeros)} \\\cline{2-6}
		& Spot  & Linear  & MOB I   & MOB II & BC  \\\hline
		BOOST-MI & 0.046(0.108) & 0.175(0.158)& \textbf{0.249}(0.166) & \textbf{0.368}(0.107) & \textbf{0.288}(0.139)  \\
		BOOST-GP & \textbf{0.085}(0.152) & \textbf{0.198} (0.146) & 0.000(0.000) & 0.070(0.145) & 0.183(0.119) \\
		BinSpect-rank & 0.046(0.122) & 0.075(0.141) & 0.021(0.100)& 0.095(0.097) & 0.197(0.102) \\
		BinSpect-km & 0.046(0.089) & 0.008(0.108) & 0.024(0.096)& 0.018(0.078) & -0.044(0.063)  \\
		SPARK & 0.000(0.000) & 0.024(0.076) & 0.000(0.000)& 0.000(0.000)& 0.048(0.101)   \\ 
		SpatialDE   & 0.000(0.000) & 0.000(0.000) & 0.000(0.000)  & 0.000(0.000)& 0.000(0.000)\\
		\hline
	\end{tabular}
	\label{mcc}
\end{table}

\clearpage

\renewcommand{\arraystretch}{0.5} 
\setlength{\tabcolsep}{6pt} 
\begin{small}
	\begin{table}[!h]
		\caption{\footnotesize Real data analysis on the mouse olfactory bulb (MOB) dataset: List of $60$ SV genes that had an attraction pattern with a positive interaction parameter in the Ising model.}
	
	\begin{tabular}{cccc}
		\hline
		
		$\quad \quad $Gene $\quad\quad$& Bayes Factor ($\text{BF}$) & $2\times \text{ln}(\text{BF})$ & Total Raw Count \\
		
		\hline
		Rc3h2 & Inf & Inf & 1174 \\
		Trib2 & Inf & Inf & 503 \\
		Med21 & Inf & Inf & 290 \\
		Nup210 & Inf & Inf & 163\\
		Rsad1 & Inf & Inf & 94\\
		Arap1 & Inf & Inf & 86\\
		Zfp938 & 9999.000 & 18.420 & 190 \\
		Trim8 & 6665.667 & 17.609 & 809\\
		Mrps18b & 2856.143 & 15.914 & 192 \\
		Rbm15b & 2499.000 & 15.647 & 151 \\
		Fam20c & 2221.222 &  15.412 & 1603 \\
		Pdik1l & 2221.222 & 15.412 & 143 \\
		Ccnh & 951.381 & 13.716 &  1261 \\
		Ercc3 & 951.381 &  13.716 & 239 \\
		Vapa & 868.565 & 13.534 & 4017\\
		Dgke & 713.286 &  13.140 & 393\\
		Zfp248 & 713.286 & 13.140 & 109 \\
		Fen1 & 644.161 & 12.936 & 88\\
		Amigo2 & 624.000 & 12.872 & 445\\
		Mfap3 & 605.061 & 12.811 & 208\\
		Katnal1 & 539.541 &  12.581 & 204 \\
		Tkt & 525.316 & 12.528 & 518 \\
		Plekhm3 & 525.316 & 12.528 & 317\\
		Ofd1 & 525.316 & 12.528 & 97 \\
		Mrps24 & 511.821 & 12.476 & 813 \\
		Wdr3 & 511.821 & 12.476 & 218 \\
		Pold3 & 499.000 & 12.425 & 227\\
		Rcbtb1 & 486.805 & 12.376 & 552\\
		Snx21 & 433.783 & 12.145 & 199\\
		Fgf14 & 407.163 & 12.018 & 158\\
		Nop58 & 376.358 & 11.861 & 248\\
		Zfp9 & 337.983 & 11.646 & 144\\
		Bbs5 & 326.869 & 11.579 & 249 \\
		Eif3c & 284.714 & 11.303 & 1123 \\
		Chpf2 & 265.667 & 11.164 & 126\\
		Ruvbl1 & 262.158 & 11.138 & 470 \\
		Zfp846 & 262.158 & 11.138 & 103\\
		Umps & 231.558 & 10.890 & 128\\

		\end{tabular}
		\label{pos}
		
	\end{table}

\begin{table}[t]
		\setlength{\tabcolsep}{15pt} 
	\begin{tabular}{cccc}
	$\ \ \ \ $& $\quad\quad\quad\quad\quad $ & $\quad\quad\quad$ & $\quad\quad\quad\quad\quad\quad\quad$ \\
		Pcnxl4 & 226.273 & 10.843 & 286\\
		Cstf2 & 214.054 & 10.732 & 780 \\
		Pias2 & 209.526 & 10.690 & 920\\
		Cog4 & 209.526 & 10.690 & 308 \\
		Dnalc1 & 205.186 & 10.648 & 616\\
		Cnih3 & 205.186 &  10.648 & 287 \\
		Phyhipl & 203.082 & 10.627 & 1488\\
		Hnrnpf & 203.082 & 10.627 & 819\\
		Ncoa2 & 195.078 & 10.547 & 953\\
		Atg13 & 195.078 & 10.547 & 728\\
		X1700025G04Rik & 187.679 & 10.469 & 794\\
		X2310003H01Rik & 184.185 & 10.432 & 94\\
		Pin1 & 175.991 & 10.341 & 913 \\
		Nceh1 & 174.439 & 10.323 & 871\\
		Lamtor5 & 172.913 & 10.306 & 356\\
		Tmem42 & 171.414 & 10.288 & 157\\
		Anks3 & 169.940 & 10.271 &  384\\
		Crot & 167.067 & 10.237 & 481\\
		Tbc1d23 & 161.602 & 10.170 & 399 \\
		Armcx2 & 157.730 & 10.122 & 594\\
		Nabp1 & 156.480 & 10.106 & 111\\
		Rasgrp2 & 154.039 & 10.074 & 170 \\
		\hline
		\end{tabular}
		
	\end{table}
\end{small}

\end{document}